\documentclass[preprint,5p,sort&compress]{myelsarticle}

\usepackage{tikz}
\usetikzlibrary{calc,positioning}
\usepackage[mode=buildnew]{standalone}
\usepackage{pgfplots}
\pgfplotsset{compat=1.12}
\usepgfplotslibrary{groupplots}
\pgfplotsset{every axis/.append style={thick}}

\usepackage{makecell}
\newcommand{\tablewordfrac}[2]{\(\displaystyle \frac{\text{#1}}{\text{#2}} \)}
\usepackage{multirow}

\usepackage[labelfont=bf]{caption}
\usepackage{amsmath}
\numberwithin{equation}{section}
\usepackage{amssymb}
\usepackage[inline]{enumitem}
\usepackage[version=4]{mhchem}
\usepackage{siunitx}
\DeclareSIUnit\Molar{\textsc{m}}
\usepackage{url}
\newsavebox{\battery}

\newcommand{\pd}[2]{\frac{\partial #1}{\partial #2}}

\newcommand{\bse}{\begin{subequations}}
\newcommand{\ese}{\end{subequations}}
\newcommand{\myint}[4]{\int_{#1}^{#2} \! #3 \, \mathrm{d}#4}
\newcommand{\bm}[1]{\mathbf{#1}}
\newcommand{\di}[1]{\hat{#1}}
\newcommand{\Deff}{D^\text{eff}}
\newcommand{\keff}{\kappa^\text{eff}}
\newcommand{\diDeff}{\di{D}^\text{eff}}
\newcommand{\dikeff}{\di{\kappa}^\text{eff}}
\newcommand{\seff}{\sigma^\text{eff}}
\newcommand{\Cd}{\mathcal{C}_\text{d}}
\newcommand{\dijecd}{\di{j}_0}
\newcommand{\jecd}{{j}_0}
\newcommand{\jecdn}{{j}_{0,\text{n}}}
\newcommand{\jecdp}{{j}_{0,\text{p}}}
\newcommand{\ocpn}{_{\ce{Pb}}}
\newcommand{\ocpp}{_{\ce{PbO_2}}}

\newcommand{\sol}{_{\text{s}}}
\newcommand{\n}{_\text{n}}

\newcommand{\p}{_\text{p}}
\newcommand{\elln}{\ell_\text{n}}

\newcommand{\ellp}{\ell_\text{p}}

\begin{document}

\title{Faster Lead-Acid Battery Simulations from Porous-Electrode Theory: \\
I. Physical Model}

\author[mi]{Valentin~Sulzer\corref{cor1}}
\ead{sulzer@maths.ox.ac.uk}
\author[mi,far]{S.~Jon~Chapman}
\ead{chapman@maths.ox.ac.uk}
\author[mi,far]{Colin~P.~Please}
\ead{please@maths.ox.ac.uk}
\author[eng,far]{David~A.~Howey}
\ead{david.howey@eng.ox.ac.uk}
\author[eng,far]{Charles~W.~Monroe}
\ead{charles.monroe@eng.ox.ac.uk}

\cortext[cor1]{Corresponding author}
\address[mi]{Mathematical Institute, University of Oxford, OX2 6GG, United Kingdom}
\address[eng]{Department of Engineering Science, University of Oxford, OX1 3PJ, United Kingdom}
\address[far]{The Faraday Institution}

\begin{abstract}
An isothermal porous-electrode model of a discharging lead-acid battery is presented, which includes an extension of concentrated-solution theory that accounts for excluded-volume effects, local pressure variation, and a detailed microscopic water balance. The approach accounts for three typically neglected physical phenomena: convection, pressure diffusion, and variation of liquid volume with state of charge. Rescaling of the governing equations uncovers a set of fundamental dimensionless parameters that control the battery's response. Total volume change during discharge and nonuniform pressure prove to be higher-order effects in cells where variations occur in just one spatial dimension. A numerical solution is developed and exploited to predict transient cell voltages and internal concentration profiles in response to a range of C-rates. The dependence of discharge capacity on C-rate deviates substantially from Peukert's simple power law: charge capacity is concentration-limited at low C-rates, and voltage-limited at high C-rates. The model is fit to experimental data, showing good agreement.
\end{abstract}
\maketitle

\section{Introduction}

Lead-acid batteries are the most widely used electrochemical storage technology, with applications including car batteries and off-grid energy supply. Models can improve battery management---for example, by minimising overcharge to extend cycle life.
% In this paper, we use concentrated-solution theory to develop a thermodynamically consistent, isothermal porous-electrode model of a discharging lead-acid battery that takes into account excluded-volume effects and Faradaic convection \cite{liu2014solute}.

The most rigorous mechanistic approach to battery-cell modelling begins with a detailed microscopic description, wherein the electrolyte and electrodes occupy discrete spatial domains; volume averaging is then performed to produce a macroscopic model \cite{richardson2012multiscale, schmuck2015homogenization, trainham2011flow}. The details are beyond the present scope, but such a homogenization underpins the model presented below.

Our development of a detailed macrohomogeneous model of a typical lead-acid battery augments standard approaches by explicitly considering the balance of water, the variation of acid density with molarity, and the distribution of pressure. As well as ensuring thermodynamic consistency and retaining model closure when systems span multiple spatial dimensions~\cite{liu2014solute}, this reveals novel convective phenomena that may occur at high discharge currents.

Nondimensionalization of the model helps to assess the relative importances of different multiphysical phenomena within it. This facilitates numerical implementation because it greatly improves the conditioning of the system, and hence improves the speed of computations, by making most dependent variables close to unity.
This paper focuses on the development of a detailed nonisobaric physical model of battery discharge. A novel nondimensionalization is presented, which helps to identify the key dimensionless parameters that control the battery's transient response. Finally, a numerical procedure is developed to solve the nonlinear system of partial differential equations comprising the model. The results are fit against experimental discharge data and used to show how discharge capacity depends on C-rate. The nonlinear model shows that the battery response does not satisfy Peukert's law. Instead, the capacity follows one power law at low C-rates, where average acid concentration controls the response, and a different power law at at high C-rates, where overpotentials control the response.

We show below that some of the excluded-volume and pressure effects can be neglected in the detailed physical model, on the basis that certain dimensionless factors are small for the particular materials used in the battery considered here. In cases where dimensionless parameters in a model are small or large, asymptotic analysis can be employed, producing simplified, more computationally efficient models that achieve the same level of physical accuracy. The full nonlinear model presented here serves as the baseline for testing a hierarchy of computationally efficient reduced-order models, which will be developed by perturbation expansion in part II.

%Upon solving our model, we identify an extended Peukert's Law: the capacity is concentration-limited at low C-rates, and voltage-limited at high C-rates.

\section{Model}
\label{sec:model}

\subsection{Unscaled governing system}
\label{sec:dim_model}
%\label{sec:chem_and_geom}

\begin{figure}
	\centering
	\includegraphics{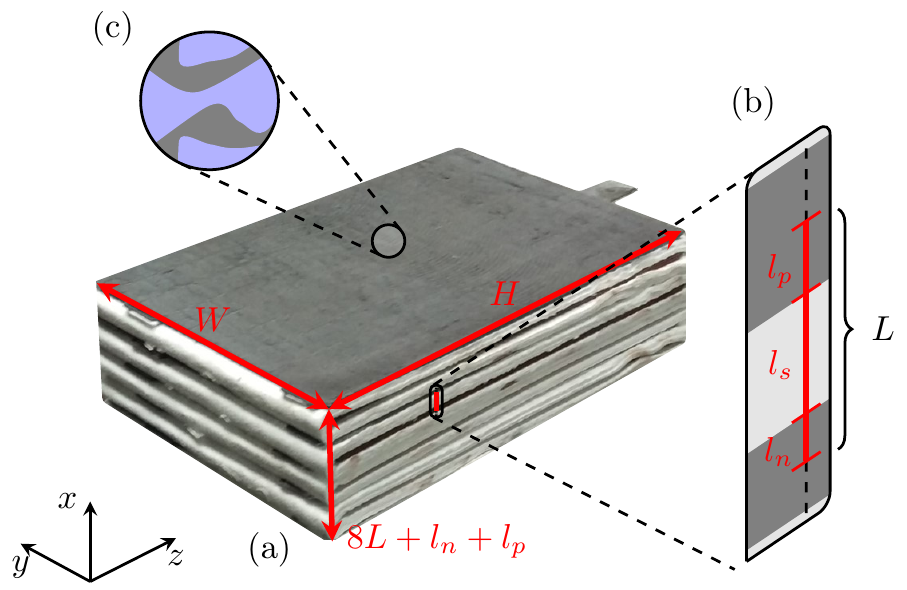}
	\caption{Geometry of a lead-acid cell. {(a)} macroscopic shape and dimensions; {(b)} (half-)widths of electrodes and separator; (c) schematic of a volume element of the porous electrode. The total width of the cell is $8L+l_\text{n}+l_\text{p}$ because $l_\text{n}$ and $l_\text{p}$ are electrode half-widths. Photograph by Ashley Grealish (BBOXX).}
	\label{fig:lead-acid_geometry}
	\end{figure}

\paragraph{Battery configuration and chemistry}
Typically, a valve regulated lead-acid battery comprises six 2 V {cells} wired in series. Figure \ref{fig:lead-acid_geometry} depicts one such {cell}, which consists of five lead ($\ce{Pb}$) electrodes and four lead dioxide ($\ce{PbO_2}$) electrodes, sandwiched alternatingly around a porous, electrically insulating separator to produce eight {electrode pairs}, wired in parallel at the top edge of the electrode pile.
The pile has height $H$, depth $W$, and cross-sectional area $A_\text{cs} = HW$. The negative ($\ce{Pb}$) and positive ($\ce{PbO_2}$) electrodes have half-widths~$l_\text{n}$ and $l_\text{p}$ respectively, and the separator has width $l_\text{sep}$, giving each internal electrode pair total width $L = {l_\text{n}+l_\text{sep}+l_\text{p}}$. The layers repeat periodically, allowing the whole pile to be modelled via analysis of one electrode pair. Some thermal effects have been documented experimentally~\cite{culpin2004thermal, vaccaro1991experiments} and considered theoretically~\cite{gu2002modeling, srinivasan2003analysis}; we assume an isothermal system for simplicity.

The electrodes are porous, permeated by an aqueous sulfuric acid ($\ce{H_2SO_4}$) electrolyte that also permeates the separator. The negative and positive half-reactions are
\begin{equation}\label{eq:halfreactions}
\begin{array}{l}
\ce{Pb + HSO_4^-
<=>[\text{\tiny discharge}][\text{\tiny charge}]
${\ce{PbSO_4 + H^+ + 2 e^-}}$
}~~\text{and} \\
\ce{PbO_2 + HSO_4^- + 3H^+ + 2e^-
<=>[\text{\tiny discharge}][\text{\tiny charge}]
${\ce{PbSO_4 + 2H_2O}}$,
}
\end{array}
\end{equation}
respectively. In both reactions, solid lead sulfate ($\ce{PbSO_4}$) forms as bisulfate anions ($\ce{HSO_4^-}$) leave the liquid.

Without competitive chemistry, reactions \eqref{eq:halfreactions} would simply reverse during recharge, as indicated. However, some authors~\cite{bernardi1995mathematical, cugnet2011effect, gu2002modeling, newman1997simulation} have observed that {gas evolution} can become important when recharging. To avoid the need for considering such side reactions, the present analysis is limited to discharges.

\paragraph{Liquid thermodynamics}
In water ($\ce{H_2O}$), $\ce{H_2SO_4}$ dissociates mainly to hydrogen cations ($\ce{H^+}$) and $\ce{HSO_4^-}$; in keeping with prior models~\cite{bernardi1995mathematical, bernardi1993two, gu1987mathematical, gu1997numerical}, speciation to sulfate is neglected. Thus the pore-filling liquid comprises $\ce{H_2O}$, $\ce{HSO_4^-}$, and $\ce{H^+}$, with partial molar volumes $\bar{V}_\text{w}$, $\bar{V}_-$ and $\bar{V}_+$, molar masses $M_\text{w}$, $M_-$, and $M_+$, and equivalent charges $0$, $-1$, and $+1$, respectively. Unscaled variables are denoted with a hat: for example, the species molarities are $\di{c}_\text{w}$, $\di{c}_-$, and $\di{c}_+$.

On any scale much larger than the Debye length, local electroneutrality holds in the liquid phase. Both ion concentrations then relate to the $\ce{H_2SO_4}$ molarity $\di{c}$ through
\begin{equation}\label{eq:electroneut}
\di{c} = {\di{c}_+} = {\di{c}_-}.
\end{equation}
In an isothermal state, thermodynamic consistency of the liquid volume requires that $\di{c}_\text{w}$ depends on $\di{c}$ alone if the bulk modulus of the liquid is very large~\cite{goyal2017new}, through
\begin{equation}\label{eq:cw}
\di{c}_\text{w} = \frac{1-\bar{V}_\text{e}\di{c}}{\bar{V}_\text{w}},
\end{equation}
where $\bar{V}_\text{e} = \bar{V}_+ + \bar{V}_-$ is the partial molar volume of the acid. Consequently the total liquid molarity $\di{c}_\text{T} = 2\di{c} + \di{c}_\text{w}$
also depends only on $\di{c}$. Acid molarity further determines the mass density $\di{\rho}$ of the liquid, because
\begin{equation}
\di{\rho} = \di{c}_\text{w} M_\text{w} + \di{c} M = \frac{M_\text{w}}{\bar{V}_\text{w}} + \left( \frac{M_\text{e}}{\bar{V}_\text{e}} - \frac{M_\text{w}}{\bar{V}_\text{w}} \right) \bar{V}_\text{e} \di{c},
\end{equation}
in which $M_\text{e} = M_+ + M_-$ stands for the acid's molar mass. Experiments show that $\di{\rho}$ varies almost linearly with $\di{c}$~\cite{chapman1968compilation}, so the partial molar volumes can be assumed constant.

The mechanical state of the liquid is described by the external pressure $\di{p}$, while its mixing free energy is parameterized by a thermodynamic Darken factor $\di{\chi} \left( \di{c} \right)$.

\paragraph{Balances}
The position $\di{x}$ within each electrode pair traverses three subdomains: the negative electrode ($0<\di{x}<l_\text{n}$), separator ($l_\text{n}<\di{x}<L-l_\text{p}$), and positive electrode ($L-l_\text{p}<\di{x}<L$). The model structure is identical in each subdomain. Parameters describing the electrolyte are similar everywhere, while those describing pore geometry generally differ among subdomains; quantities that parameterise reactions differ in the negative- and positive-electrode subdomains, and vanish in the separator.

Liquid volume fraction $\di{\varepsilon}$ and reactive solid area per volume $\mathcal{A}$ characterize the homogenized geometry within the electrode pair. In electrode subdomains, deposition (removal) of solid $\ce{PbSO_4}$ on pore surfaces is accompanied by removal (deposition) of solid $\ce{Pb}$ or $\ce{PbO_2}$, so these geometric parameters can generally vary locally. Solid $\ce{PbSO_4}$ does not tend to deposit as a compact thin film, so mechanistic lead-acid battery models usually let $\mathcal{A}$ vary with state of charge~\cite{bernardi1995mathematical, bernardi1993two, gu1987mathematical,
gu1997numerical, newman1997simulation}.
Since the functionality of this variation is disputed, we instead let the area be constant, following the approach of Liu \textit{et al.}, who showed this is sufficient to model Li/air-battery discharge~\cite{liu2016capacity}. From this perspective $\mathcal{A}$ describes an immobile Gibbs dividing surface that partitions the layers of Pb or $\ce{PbO_2}$ that contribute to the battery's charge state from the current-collecting Pb layer beneath them, which does not. Thus the time change in pore volume relates to the molar volumes of the solid species in scheme \eqref{eq:halfreactions} through
\begin{equation}\label{eq:dim_depsdt}
\frac{1}{\mathcal{A}} \pd{\di{\varepsilon}}{\di{t}} = -
\bar{V}_{\text{PbSO}_4} \di{S}_{\text{PbSO}_4} - \bar{V}_{\text{Pb}} \di{S}_{\text{Pb}} - \bar{V}_{\text{PbO}_2} \di{S}_{\text{PbO}_2},
\end{equation}
in which $\bar{V}_k$ is the partial molar volume of species $k$ and $\di{S}_k$ is the rate at which $k$ is generated by interfacial reactions per unit of pore area.

After homogenization, the local mass balance of species
$k \in \left\{ \text{w}, -, + \right\}$
in the pore-filling liquid phase implies that
\begin{equation}\label{eq:dim_cons_c}
\pd{}{\di{t}}(\di{\varepsilon} \di{c}_k) = -\di{\nabla}\cdot\di{\bm{N}}_k + \mathcal{A} \di{S}_k,
\end{equation}
where $\di{\bm{N}}_k$ represents the molar flux of $k$. With electroneutrality condition \eqref{eq:electroneut} and equation of state \eqref{eq:cw}, all three balances \eqref{eq:dim_cons_c} combine to show liquid-volume continuity,
\begin{equation}\label{eq:dim_vbox0}
\frac{\partial \di{\varepsilon} }{\partial \di{t} } = - \di{\nabla}\cdot\di{\bm{v}}^\square + \mathcal{A} \left( \bar{V}_\text{w} \di{S}_\text{w} +\bar{V}_- \di{S}_- +\bar{V}_+ \di{S}_+ \right),
\end{equation}
in which $\di{\bm{v}}^\square = \bar{V}_\text{w} \di{\bm{N}}_\text{w} + \bar{V}_- \di{\bm{N}}_- +\bar{V}_+ \di{\bm{N}}_+$ signifies the volume-averaged liquid velocity.

Under Faraday's law, ion balances \eqref{eq:dim_cons_c} also combine to demonstrate charge continuity in the liquid,
\begin{equation}\label{eq:dim_divi}
\di{\nabla}\cdot\di{\bm{i}} = \mathcal{A}\di{j}.
\end{equation}
Letting $F$ be Faraday's constant, the liquid-phase current density is $\di{\bm{i}} = F ( \di{\bm{N}}_+ - \di{\bm{N}}_- )$ and the current density associated with interfacial charge exchange is $\di{j} = F ( \di{S}_+ - \di{S}_- )$. (Positive $\di{j}$ flows into pores.) Note that any current leaving the liquid at a given location enters the solid there. Thus
\begin{equation}\label{eq:dim_divis}
\di{\nabla}\cdot\di{\bm{i}}_{\text{s}} = -\mathcal{A} \di{j},
\end{equation}
where $\di{\bm{i}}_\text{s}$ is the solid-phase current density.

In a general isothermal setting, liquid convection is determined by a momentum balance, such as Cauchy's equation~\cite{liu2014solute,goyal2017new}. This governs the distribution of momentum density $\di{\rho} \di{\bm{v}}$, naturally expressed in terms of the mass-averaged velocity $\di{\bm{v}}$. The kinematic relation
\begin{multline}\label{eq:dim_kinematic}
\di{\bm{v}} - \di{\bm{v}}^\square = \frac{\bar{V}_\text{e} }{\di{\rho}} \left( \frac{M_\text{e} }{\bar{V}_\text{e} } - \frac{M_\text{w} }{\bar{V}_\text{w} } \right) \left( \di{\bm{N}}_+ - \di{c} \di{\bm{v}}^\square \right)
\\ + \frac{\bar{V}_-}{\di{\rho}} \left( \frac{M_\text{w} }{\bar{V}_\text{w}} - \frac{M_- }{\bar{V}_-} \right) \frac{\di{\bm{i}} }{F},
\end{multline}
specifies how $\di{\bm{v}}$ must relate to $\di{\bm{v}}^\square$.

\paragraph{Flux constitutive laws}
Two Onsager--Stefan--Maxwell flux laws govern transport in the liquid phase. The law for the thermodynamic force on water~\cite{goyal2017new} can be inverted \cite{monroe2013continuum} to give
\begin{equation}\label{eq:dim_N+}
\di{\bm{N}}_+
= -\diDeff \left(\di{\nabla} \di{c} -\frac{\di{\psi}\di{\nabla} \di{p}}{RT \di{\chi}} \right) + \frac{t_+^\text{w} \di{\bm{i}} }{F} + \di{c} \di{\bm{v}}^\square,
\end{equation}
where $R$ is the gas constant, $T$ is the absolute temperature, $t^\text{w}_+$ is the cation transference number relative to the water velocity, and $\diDeff$ is the effective diffusivity of acid in water; the pressure-diffusion factor $\di{\psi}$ depends on $\di{c}$ through
\begin{equation}
\di{\psi} = \frac{\bar{V}_\text{w} \di{c}_\text{T} M_\text{w} \di{c}_\text{w} M_\text{e} \di{c} }{2 \di{\rho} } \left( \frac{\bar{V}_\text{w}}{M_\text{w}} - \frac{\bar{V}_\text{e} }{M_\text{e} } \right).
\end{equation}
(To put equation \eqref{eq:dim_N+} in the form given, one must have that $\bar{V}_+ = ( 1 - t^\text{w}_+)\bar{V}_\text{e}$ and $\bar{V}_- = t^\text{w}_+ \bar{V}_\text{e}$~\cite{newman1973restricted}.)
The law for the thermodynamic force on cations can be linearly transformed to produce a current/voltage relation:
\begin{multline}\label{eq:dim_i}
\di{\bm{i}} = - \dikeff \di{\nabla} \di{\Phi} + \frac{2RT\di{\chi}(1-t^\text{w}_+) \dikeff}{F\bar{V}_\text{w} \di{c}_\text{T}}\di{\nabla} \ln \di{c} \\
+ \frac{\dikeff}{F \di{c} } \left[ \frac{M_+ \di{c} }{\di{\rho} } - \frac{2 \di{\psi} (1-t^\text{w}_+) }{ \bar{V}_\text{w} \di{c}_\text{T} } \right] \di{\nabla} \di{p}.
\end{multline}
Here $\dikeff$ is the effective conductivity; $\di{\Phi}$ is the potential measured by a reversible hydrogen electrode at $\di{p}$~\cite{bizeray2016resolving}. Effective properties appear in equations \eqref{eq:dim_N+} and \eqref{eq:dim_i} because pore connectivity affects apparent transport rates. We let
\begin{equation}
\diDeff = \di{D}(\di{c})\di{\varepsilon}^{3/2}~~~\text{and}~~~\dikeff = \di{\kappa}(\di{c})\di{\varepsilon}^{3/2},
\end{equation}
following Bruggeman's tortuosity correlation.

The solid phase is electronically but not ionically conductive. Thus the current density there relates to the solid-phase potential, $\di{\Phi}_\text{s}$, through Ohm's law,
\begin{equation}
\di{\bm{i}}_{\text{s}} = -\seff \di{\nabla}\di{\Phi}_{\text{s}},
\end{equation}
where $\seff$ is the effective conductivity,
\begin{equation}
\seff = \sigma \di{\varepsilon}_{\text{s}}^{3/2}.
\end{equation}
Here $\seff$ is constant because $\di{\varepsilon}_\text{s}$, the volume fraction of nonreacting Pb bounded within electron-exchange surface $\mathcal{A}$, does not vary. Note that $\di{\varepsilon}_\text{s}$ does not count the volume fraction occupied by solid reactants, so $\di{\varepsilon}_\text{s} \neq 1-\di{\varepsilon}$.

A constitutive law for liquid stress closes the model \cite{liu2014solute}. Below, terms involving momentum will be shown to be negligible in the first approximation. To analyse the scale of these terms, it will suffice to assume that current density induces a flow with low Reynolds number, in which case the homogenization of Cauchy's equation produces Darcy's law,
\begin{equation}\label{eq:dim_Darcy}
\di{\bm{v}} = -\frac{\di{\mathcal{K}}}{\di{\mu}}\di{\nabla}\di{p},
\end{equation}
in which $\di{\mu}$ is the liquid viscosity; pore geometry determines the Kozeny--Carman permeability $\di{\mathcal{K}}$.

\paragraph{Interfacial constitutive laws}
Electrons are the only solid-phase charge carrier in scheme \eqref{eq:halfreactions}, making interfacial electronic current a proxy for the half-reaction rates. No interfacial reactions occur in the separator domain, so $\di{j} = 0$ uniformly there and the reactive area $\mathcal{A}$ does not need to be defined there.

In an electrode subdomain where a single half-reaction involving $n_\text{e}$ electrons occurs, the molar flux of reactants is $\di{j} / n_\text{e} F$ and hence every $\di{S}_k$ is given by
\begin{equation}\label{eq:dim_S+}
\di{S}_k = - \frac{s_k \di{j}}{n_\text{e} F},
\end{equation}
where $s_k$ is the signed stoichiometric coefficient of species $k$ in the half-reaction. (For a reduction half-reaction, $s_k$ is positive for a product and negative for a reactant.)

Relationship \eqref{eq:dim_S+} permits simpler expressions to be used in place of the general material balances \eqref{eq:dim_cons_c}. First, equations \eqref{eq:dim_depsdt}, \eqref{eq:dim_cons_c}, \eqref{eq:dim_vbox0} and \eqref{eq:dim_S+} combine to give
\begin{equation}\label{eq:dim_vbox}
\di{\nabla}\cdot\di{\bm{v}}^\square = - \frac{ \Delta \bar{V} \mathcal{A} j }{F} .
\end{equation}
This liquid-phase volume balance introduces the volume of reaction $\Delta \bar{V}$, related to half-reaction stoichiometry by
\begin{multline}
n_\text{e} \Delta \bar{V} = \bar{V}_\text{w} s_\text{w} + \bar{V}_- s_- + \bar{V}_+ s_+ + \bar{V}_{\text{PbSO}_4} s_{\text{PbSO}_4} \\
+ \bar{V}_{\text{Pb}} s_{\text{Pb} }  + \bar{V}_{\text{PbO}_2} s_{\text{PbO}_2 }.
\end{multline}
Second, one can combine equations \eqref{eq:electroneut}, \eqref{eq:dim_depsdt}, \eqref{eq:dim_cons_c}, \eqref{eq:dim_divi}, \eqref{eq:dim_N+}, \eqref{eq:dim_S+}, and \eqref{eq:dim_vbox} to show that the acid is governed by a form of the convective diffusion equation:
\begin{multline}\label{eq:dim_saltbal}
\pd{}{\di{t}}\left(\di{\varepsilon}  \di{c} \right) + \di{\bm{v}}^\square \cdot \di{\nabla} \di{c} = \di{\nabla} \cdot \left[ D^{\text{eff} } \left( \di{\nabla} \di{c} - \frac{\di{\psi} \di{\nabla} \di{p} }{\di{\chi} RT} \right) \right] \\
+ \left( s + \di{c} \Delta \bar{V} \right) \frac{\mathcal{A} \di{j} }{F} .
\end{multline}
The generation term here includes a single additional parameter,
\begin{equation}
s = -\frac{s_+ + n_\text{e} t_+^\text{w} }{n_\text{e}}.
\end{equation}
Three aspects of the acid balance \eqref{eq:dim_saltbal} are new. First, the convection term, and second, the volume of reaction $\Delta \bar{V}$, appear because state equation \eqref{eq:cw} imposes constraints on balances \eqref{eq:dim_cons_c}. Finally, a pressure-diffusion term appears because the flux laws are based on thermodynamic forces.

To complete the model, the current density $\di{j}$ across the surface $\mathcal{A}$ is governed by a chemical-kinetic constitutive law. Generally such laws involve the voltage difference between the liquid and solid, the equilibrium potential of the half-reaction, and the chemical activities of the reactants. We assume the half-reactions in scheme \eqref{eq:halfreactions} are elementary, following Butler--Volmer kinetics.
Butler--Volmer laws naturally include a symmetry factor, which we take to be one half~\cite{gu1987mathematical}, yielding
\begin{equation}\label{eq:dim_Butler-Volmer}
\di{j} = 2 \dijecd\sinh \left( \frac{F \di{\eta} }{RT} \right)
+ C_{\text{dl}} \frac{\partial ( \di{\Phi}_\text{s} - \di{\Phi} ) }{\partial \di{t} } ,
\end{equation}
where $\dijecd$ is the concentration-dependent exchange-current density,
\begin{equation}\label{eq:dim_j0}
\dijecd = j^{\text{ref}} \left( \frac{\di{c} }{c^{\text{ref}} } \right)^{ \left| \frac{s_+}{n_{\text{e}}} \right| + \left| \frac{s_-}{n_{\text{e}}} \right| } \left( \frac{\di{c}_{\text{w}} }{c_{\text{w}}^{\text{ref}} } \right)^{\left| \frac{s_\text{w}}{n_{\text{e}}} \right| } ,
\end{equation}
and $\di{\eta}$ is the surface overpotential
\begin{equation}\label{eq:eta}
\di{\eta} = \di{\Phi}_{\text{s}} - \di{\Phi} - \di{U}.
\end{equation}
Here $\di{U}$ stands for the half-reaction's open-circuit potential (OCP) relative to a particular reference electrode. (The reference must be the same for all half-reactions.) Terms involving interfacial capacitance $C_{\textrm{dl}}$ help to smooth out numerics, but have a negligible effect on model predictions because the capacitive time-scale is very short~\cite{srinivasan2003mathematical}.

\paragraph{Boundary conditions}
Equations \eqref{eq:dim_depsdt}, \eqref{eq:dim_divi} to \eqref{eq:dim_kinematic}, and \eqref{eq:dim_i} to \eqref{eq:eta} comprise a three-dimensional model with closure at every interior point in an electrode pair.
Symmetry and insulating boundary conditions demand that no species in the liquid phase flows through the centers, sides, and bottom of the electrode pair, so
\begin{multline}\label{eq:no-flux_sides}
\di{\bm{N}}_+\cdot\bm{n} = \di{\bm{i}}\cdot\bm{n} = \di{\bm{v}}^\square\cdot\bm{n} = 0 \\ \text{ at $\di{x}=0,L$, $\di{y}=0,W$ and $\di{z}=0$.}
\end{multline}
Kinematic relation \eqref{eq:dim_kinematic} shows that $\di{\bm{v}}\cdot\bm{n} = 0$ at these boundaries too. Flux laws \eqref{eq:dim_N+}, \eqref{eq:dim_i}, and \eqref{eq:dim_Darcy} further require the surface-normal gradients
$\partial \di{c}/\partial n$, $\partial \di{p}/\partial n$ and $\partial \di{\Phi}/\partial n$
to vanish at these boundaries.

Above the electrodes, there is a region of free electrolyte, with height $\di{h}(\di{t})$. At the top surface of this region, we impose a known external pressure $\di{p}_\text{ext}$, and the absence of flux relative to the surface, which moves with velocity ${\bm{\di{v}}^\text{head} = \left(\partial\di{h}/\partial\di{t}\right)\bm{e}_z}$:
\begin{multline}\label{eq:top_bcs}
\left(\bm{\di{N}}_+-\bm{\di{v}}^\text{head}\di{c}\right)\cdot\bm{n}=\bm{\di{i}}\cdot\bm{n}=\left(\bm{\di{v}^\square}-\bm{\di{v}}^\text{head}\right)\cdot\bm{n}=0, \\\di{p} = \di{p}_\text{ext}, \quad \text{ at $\di{z} = H+\di{h}(\di{t})$}.
\end{multline}
Note the final condition in \eqref{eq:top_bcs} determines the \textit{a priori} unknown height $\di{h}(\di{t})$.

Hereafter, subscripts $\text{n}$ and $\text{p}$ denote property values in the negative- and positive-electrode subdomains. We choose the negative electrode to be the ground state, and define the cell voltage to be the potential at the positive electrode tab:
\begin{equation}
\di{\Phi}_{\text{s}}
= \begin{cases}
0 &\quad \hat{\bm{x}} \in \text{tab}\n, \\
\di{V}(\di{t}) &\quad \hat{\bm{x}} \in \text{tab}\p.
\end{cases}
\end{equation}
One can either control the voltage, or consider a current-controlled discharge where the voltage is determined by
\begin{equation}\label{eq:is_BC}
- \myint{\text{tab}\n}{}{\di{\bm{i}}\sol\cdot\bm{n}}{S}
= \myint{\text{tab}\p}{}{\di{\bm{i}}\sol\cdot\bm{n}}{S}
= \mathrm{\di{I}}_\text{circuit}(\di{t})/8,
\end{equation}
where $\di{\mathrm{I}}_\text{circuit}$ is the current drawn from the battery, which is positive for a discharge; the factor of 8 appears because the cell comprises eight electrode pairs in parallel.

This paper focuses on experiments under `galvanic control', following condition \eqref{eq:is_BC},
which allow $\di{\mathrm{I}}_\text{circuit} \left( \di{t} \right)$ to be any function of time.
Since six cells are connected in series, the voltage in the external circuit is
$\di{V}_\text{circuit}
=6\di{V}$.

\paragraph{Relationships between subdomains}
The liquid phase permeates all three subdomains. Therefore scalar variables $\di{c}$, $\di{p}$, and $\di{\Phi}$, as well as the normal components of all flux vectors,
are continuous across electrode/separator boundaries.

There is no solid-phase current at either edge of the separator subdomain or pore-surface charge exchange within it, so $\di{{i}}_\text{s}$ vanishes uniformly there. Since the separator subdomain electronically isolates the positive and negative electrodes, $\di{\Phi}_\text{s}$ is not continuous across it.

Integrating the interfacial current distributions in equation \eqref{eq:dim_divis} and applying the divergence theorem, boundary conditions \eqref{eq:is_BC} and the fact that $\di{{i}}_\text{s}$ vanishes at the electrode/separator interfaces leads to integral constraints,
\begin{multline}\label{eq:dim_j_BC}
\myint{0}{H}{
	\!\myint{0}{W}{
		\!\myint{0}{l_\text{n}}{\mathcal{A}_\text{n}\di{j}_\text{n}}{\di{x}}
	}{\di{y}}
}{\di{z}} \\
= -\myint{0}{H}{
	\!\myint{0}{W}{
		\!\myint{L-l_\text{p}}{L}{\mathcal{A}_\text{p}\di{j}_\text{p}}{\di{x}}
	}{\di{y}}
}{\di{z}}
= \di{\mathrm{I}}_\text{circuit}/8.
\end{multline}
In short, these say that the total current leaving the negative electrode domain must enter the positive electrode domain. Expressions~\eqref{eq:dim_j_BC} will help to analyze the scales of pressure and velocity in \eqref{eq:v_p_scales}.

\paragraph{Initial conditions}
The initial electrolyte concentration and electrode porosities are spatially uniform, but depend on the state of charge, $q$, which we define as
\begin{equation}\label{eq:SOC_defn}
q = q^0 - \frac{1}{q^\text{max}}\myint{0}{\di{t}}{\mathrm{I}_\text{circuit}(s)}{s}.
\end{equation}
Let $c^\text{max}$, $\varepsilon^\text{max}_\text{n}$ and $\varepsilon^\text{max}_\text{p}$ be the values of electrolyte concentration, negative electrode porosity, and positive electrode porosity, respectively, at full state of charge.
In a lead-acid battery, the state of charge is closely linked to the concentration of the electrolyte. Hence $q$ is chosen to be unity when the concentration of the electrolyte is at its maximum value, $c^\text{max}$, and zero when the concentration of the electrolyte is zero, so that the initial conditions are
\begin{equation}\label{eq:dim_ICs}
\di{c} = \di{c}^0 = c^\text{max}q^0, \quad \di{\varepsilon} = \di{\varepsilon}^0 = \varepsilon^\text{max}-\varepsilon^\Delta(1-q^0) \quad \text{ at $\di{t}=0$}.
\end{equation}
Parameters $q^\text{max}$ and $\varepsilon^\Delta$ are chosen to make \eqref{eq:SOC_defn} and \eqref{eq:dim_ICs} consistent with \eqref{eq:dim_saltbal} and \eqref{eq:dim_depsdt} (see \ref{app:params} for details):
\begin{align}
q^\text{max} &= \frac{8FA_\text{cs}\left(l_\text{n}\varepsilon_\text{n}^\text{max} + l_\text{sep}\varepsilon_\text{sep}^\text{max} + l_\text{p}\varepsilon_\text{p}^\text{max}\right) c^\text{max}}{s_\text{p}-s_\text{n}}, \label{eq:Qmax}\\
\varepsilon^\Delta &= \frac{\Delta\bar{V}^\text{surf}q^\text{max}}{16FA_\text{cs}l}.%, \quad \text{i} = \text{n},\text{p}.
\end{align}
Finally, because interfacial capacitance effects are included, initial conditions are needed for the potentials; we choose spatially homogeneous values such that $\di{j}=0$ at $\di{t}=0$:
\bse
\begin{align}
&\di{\Phi} = - \di{U}\ocpn(\di{c}^0), \\
&\di{\Phi}_{\text{s}} = \begin{cases}
0, \quad &0<\di{x}<l_\text{n}\\
\di{U}\ocpp(\di{c}^0) - \di{U}\ocpn(\di{c}^0), \quad &L-l_\text{p}<\di{x}<L.
\end{cases}
\end{align}
\ese

\subsection{Dimensionless model}
\label{sec:nondim}

Nondimensionalization of the system presented in the Unscaled governing system section
helps to determine the dominant effects in the system.
If $\partial \di{h}/\partial \di{t}$ is sufficiently small and the electrode conductivity is sufficiently high, %\footnote{The parameter regimes that ensure these assumptions are valid will not be discussed here. We return to them in \ref{app:velocity}.},
one can assume uniformity in the plane normal to $\di{x}$, reducing the problem to one spatial dimension. In this case, the tabs cover the whole electrodes at $\di{x}=0$ and $\di{x}=L$, and so the boundary condition \eqref{eq:is_BC} becomes
\begin{equation}\label{eq:is_BC_1D}
\di{\bm{i}}\sol\cdot\bm{e}_x
= \mathrm{\di{i}}_\text{cell}(\di{t})
= {\mathrm{\di{I}}_\text{circuit}}/{8A_\text{cs}} \quad \text{ at } \di{x} = 0, L,
\end{equation}
which introduces the variable $\di{i}_\text{cell}$ to stand for the total current density at the cell level.
Let $\di{N}_+$, $\di{i}$, $\di{i}_\text{s}$, $\di{v}^\square$ and $\di{v}$ represent the $\di{x}$-components of vectors $\di{\bm{N}}_+$, $\di{\bm{i}}$, $\di{\bm{i}}_\text{s}$, $\di{\bm{v}}^\square$ and $\di{\bm{v}}$. A first set of dimensionless variables is formed by the natural rescalings
\bse\label{eq:nondimensionalization_scales}
\begin{equation}\label{eq:main_nondimensionalization_scales}
{{x}} = \di{{x}}/L, \quad
({{i}},{{i}}_\text{s}) = (\di{{i}}, \di{{i}}_\text{s})/\bar{i}, \quad
c = \di{c}/c^\text{max}, \quad
{\varepsilon} = \di{\varepsilon},
\end{equation}
where $\bar{i} = \max\limits_{\di{t}}\left(\di{\mathrm{i}}_\text{cell}(\di{t})\right)$.

Reduce the number of parameters by identifying the scale of interfacial current density, $\bar{i}/\mathcal{A}L$, and discharge time-scale, $c^\text{max}FL/\bar{i}$. Hence nondimensionalize interfacial current density, exchange-current density, and time as
\begin{align}
(j_\text{n}, \jecdn) &= (\di{j}_\text{n}, \di{j}_{0,\text{n}}) \left/\frac{\bar{i}}{\mathcal{A}_\text{n}L}\right., \\
(j_\text{p}, \jecdp) &= (\di{j}_\text{p}, \di{j}_{0,\text{p}}) \left/\frac{\bar{i}}{\mathcal{A}_\text{p}L}\right., \\
t &= \di{t}\left/\frac{c^\text{max}FL}{\bar{i}}\right..
\end{align}
\ese
After nondimensionalization, $\jecdn$ and $\jecdp$ are $\mathcal{O}(1)$ functions.
To nondimensionalize the potentials, note that the dominant exponents in the hyperbolic-sine terms of \eqref{eq:dim_Butler-Volmer} should be $\mathcal{O}(1)$, since the terms multiplying the {sinh} functions are all $\mathcal{O}(1)$. Equation \eqref{eq:OCPs_split} can be exploited to define the dimensionless open-circuit potentials
\bse
\begin{align}
{U}\ocpn(c) &= \frac{F}{RT}\left(\di{U}\ocpn(\di{c}) - U^\ominus\ocpn\right)~~~\textrm{and} \\
{U}\ocpp(c) &= \frac{F}{RT}\left(\di{U}\ocpp(\di{c}) - U^\ominus\ocpp\right).
\end{align}
\ese
Then, to ensure that the exponents in equation \eqref{eq:dim_Butler-Volmer} are $\mathcal{O}(1)$ and noting that $\di{\Phi}_\text{s} = 0$ at $\di{x} = 0$, introduce the dimensionless potentials
\bse
\begin{align}\label{eq:potential_nondimensionalization_scales}
\Phi &= \frac{F}{RT}\left(\di{\Phi} + U\ocpn^\ominus\right), \\
{\Phi}_{\text{s}} &= \begin{cases}
\frac{F}{RT}\di{\Phi}_{\text{s}}, \quad &0<\di{x}<l_\text{n}\\
\frac{F}{RT}\left(\di{\Phi}_{\text{s}} - U\ocpp^\ominus + U\ocpn^\ominus\right), \quad &L-l_\text{p}<\di{x}<L.
\end{cases}
\end{align}
\ese
Quantities $\di{D}$ (and hence $\diDeff$) and $\di{c}_\text{w}$ are appropriately scaled with their values at $c=c^\text{max}$:
\begin{align}
D(c) = \di{D}(\di{c})/D^\text{max}, \quad
c_\text{w}(c) = \di{c}_\text{w}(\di{c})/c_\text{w}^\text{max},
\end{align}
where $D^\text{max} = \di{D}(c^\text{max})$ and $c_\text{w}^\text{max} = \di{c}_\text{w}(c^\text{max})$. The conductivity and Darken thermodynamic factor rescale as
\begin{align}
&\kappa(c) = \frac{RT\hat{\kappa}(\hat{c})}{F^2{D}^\text{max}{c}^\text{max}},\quad \chi(c) = \frac{
	2(1-t^\text{w}_+)\hat{\chi}(\hat{c})
}{
	1 -\alpha c
},
\end{align}
where the quantity ${\alpha = -(2\bar{V}_\text{w} - \bar{V}_\text{e})c^\text{max}}$ is defined by Liu \textit{et al.}~\cite{liu2014solute} as the excluded-volume number.

In equation \eqref{eq:dim_j0} for the exchange-current density, the reference concentrations are taken to be ${\di{c}^\text{ref} = \di{c}^\text{max}}$ and ${\di{c}_{\text{w}}^{\text{ref}} = \di{c}_\text{w}^\text{max}}$.
The dimensionless widths of the negative electrode, separator and positive electrode become $\ell_\text{n} = l_\text{n}/L$, $\ell_\text{sep} = l_\text{sep}/L$, and $\ell_\text{p} = l_\text{p}/L$, respectively.

In the velocity equations \eqref{eq:dim_vbox}, \eqref{eq:dim_Darcy} and \eqref{eq:dim_kinematic}, $\di{\rho}$ and $\di{\mu}$ are scaled with their values at $c^\text{max}$,
and $\di{\mathcal{K}}$, with $d^2$, where $d$ is the characteristic pore size. The reaction velocity scale is $v^\text{rxn} = \bar{i}/c^\text{max}F$, and the Darcy pressure scale $\mu^\text{max}v^\text{rxn}L/d^2$, so that
\bse
\begin{align}\label{eq:v_p_scales}
\left(v^\square,v\right) &= \left(\di{v}^\square, \di{v}\right)\left/\frac{\bar{i}}{c^\text{max}F}\right., \\
\quad p &= \left(\di{p}-p^\text{ref}\right)\left/\frac{\mu^\text{max}v^\text{rxn}L}{d^2}\right.,
\end{align}
\ese
where $p^\text{ref}$ is a reference pressure (e.g. atmospheric).
This scaling transforms the equations governing $v^\square$, $v$ and $p$ to
\bse
\begin{align}
\pd{v^\square}{x} &= \beta j, \label{eq:vbox}\\
v &= -\frac{\mathcal{K}}{\mu}\pd{p}{x}, \label{eq:Darcy} \\
{\rho}\left({{v}} - {{v}}^\square\right) &= - \frac{\omega_c}{\Cd}\Deff\pd{c}{x}+ \omega_ii,\label{eq:kinematic}
\end{align}
\ese
in which the dimensionless parameters $\beta$, $\omega_c$ and $\omega_i$ are defined in Table \ref{tab:dimless_params}. Further, the effect of pressure gradients in equations \eqref{eq:dim_i} and \eqref{eq:dim_saltbal} is smaller than the effect of concentration gradients (ignoring the $\mathcal{O}(1)$ functions of concentration $\chi$ and $\psi$) by a factor of
\begin{equation}
\pi_\text{os} = \frac{\mu^\text{max}v^\text{rxn}L}{d^2RTc^\text{max}}.
\end{equation}
Since $\beta$ takes different values in the two electrodes, equation \eqref{eq:vbox} does not admit a solution where $v^\square$ vanishes at both electrode centers. However, as can be seen in Table \ref{tab:dimless_params}, the dimensionless parameters $\beta$ and $\pi_\text{os}$ are small. We assume a limit where both of these parameters are zero (and hence $\partial \di{h}/\partial \di{t} = 0$), in which case $v^\square$ vanishes everywhere, and $v$ and $p$ decouple from the other variables. The full model can then be solved to find the voltage without needing the velocity and pressure.\footnote{The limit of finite $\beta$ will be explored in a future, two-dimensional, model.}

Having decoupled flow velocity and pressure from the rest of the model, the following dimensionless system for $c$, $j$, $\varepsilon$, ${i}$, $\Phi$, ${i}_\text{s}$ and $\Phi_\text{s}$ results:
\bse\label{eq:summary}
\begin{align}
\pd{}{t}(\varepsilon c) &= \frac{1}{\Cd}\pd{}{x}\left(\Deff\pd{c}{x}\right) + sj, \label{eq:dcdt}\\
\pd{\varepsilon}{t} &= -\beta^\text{surf}j, \label{eq:depsdt}\\
\pd{i}{x} &= j, \label{eq:didx}\\
\Cd\,{i} &= \keff\left(\chi\pd{\ln(c)}{x} - \pd{\Phi}{x}\right), \label{eq:i}\\
\pd{i_{\text{s}}}{x} &= -j, \label{eq:disdx}\\
{i}_{\text{s}} &= -\iota_{\text{s}}\pd{\Phi_{\text{s}}}{x}, \label{eq:is}\\
j &= 2\jecd\sinh\left(\Phi_\text{s}-\Phi-U(c)\right) + \gamma_{\text{dl}}\pd{}{t}\left(\Phi_\text{s}-\Phi\right), \label{eq:j}
\end{align}
with boundary conditions
\begin{align}
\Phi_{\text{s}} = \pd{c}{x} = i = 0, \quad i_\text{s} = \mathrm{i}_\text{cell} \quad &\text{ at } x = 0, 1, \label{eq:BCs_collectors}\\
i_\text{s} = 0 \quad &\text{ at } x = \ell_\text{n}, 1-\ell_\text{p}.\label{eq:BCs_separator}
\end{align}
and initial conditions
\begin{align}\label{eq:ICs}
c &= q^0, \\
{\varepsilon} &= \varepsilon^\text{max}-\varepsilon^\Delta(1-q^0), \\
{\Phi} &= - {U}\ocpn(q^0), \\
{\Phi}_{\text{s}} &= \begin{cases}
0, \quad &0<{x}<\ell_\text{n}\\
{U}\ocpp\left({c}^0\right) - {U}\ocpn\left({c}^0\right), \quad &1-\ell_\text{p}<{x}<1.
\end{cases}
\end{align}
Integral condition \eqref{eq:dim_j_BC} nondimensionalizes to
\begin{equation}\label{eq:j_bc}
	\myint{0}{\elln}{{j}_\text{n}}{\di{x}}
	= -\myint{1-\ellp}{1}{{j}_\text{p}}{\di{x}}
	= i_\text{cell}.
\end{equation}
\ese
Composition dependences of the properties $D^\textrm{eff}$, $\chi$, $\kappa^\textrm{eff}$, $j_0$, $U_{\textrm{Pb}}$, and $U_{\textrm{PbO}_2}$ are established through the functions listed in table \ref{tab:dim_functions}. The four dominant dimensionless parameters,
$\Cd$, $\iota_\text{s}$, $\beta^\text{surf}$ and $\gamma_\text{dl}$, are defined in Table \ref{tab:dimless_params}, which also states physical interpretations and typical values. In particular, the diffusional C-rate can be written as
\begin{equation}
\Cd ={L^2/D^\text{max}}\times\frac{Q}{8A_\text{cs}c^\text{max}FL}\times\frac{8A_\text{cs}\bar{i}}{Q},
\end{equation}
where ${8A_\text{cs}c^\text{max}FL}$ is the volumetric capacity of the battery (in Ah), $Q$ is the capacity of the battery (in Ah), and ${\mathcal{C} = 8A_\text{cs}\bar{i}/Q}$ is the C-rate of operation.
Alternatively, one can identify $\Cd$ to be the ratio between the applied current scale, $\bar{i}$, and the scale of the liquid-phase limiting current, $i_\text{L} = c^\text{max}D^\text{max}F/L$.

In the Results section, $q^0$ will be taken to be unity (the battery starts from a fully charged state) unless explicitly stated.

\renewcommand{\arraystretch}{2.5}
\begin{table*}[t]
\centering
\begin{tabular}{|c|c|c|c c c|}
\hline
\multirow{2}{*}{Parameter} & \multirow{2}{*}{Definition} & \multirow{2}{*}{Interpretation} & \multicolumn{3}{c|}{Value} \\
\cline{4-6}
& & & n & sep & p \\
\hline
$\Cd$ & \(\displaystyle \frac{\bar{i}}{c^\text{max}D^\text{max}F/L} \) & \tablewordfrac{applied current density}{limiting current density} & \multicolumn{3}{c|}{$0.60\mathcal{C}$} \\
$\iota_\text{s}$ & \(\displaystyle \frac{\seff RT/FL}{\bar{i}} \) & \tablewordfrac{Ohmic current scale}{applied current density} & $3.8\times10^4/\mathcal{C}$ & - & $55/\mathcal{C}$ \\
$\beta^\text{surf}$ & \(\displaystyle -\frac{c^\text{max}}{n_\text{e}}\sum_{k\in\{\ce{PbSO_4}, \ce{Pb}, \ce{PbO_2}\}}\bar{V}_ks_k \) & \makecell{Change in porosity \\ associated with local half-reaction \\ going to completion} &  $0.084$ & - & $-0.064$ \\
$\beta$ & \(\displaystyle -c^\text{max}\Delta \bar{V} \) & \makecell{Change in acid volume fraction \\ associated with local half-reaction \\ going to completion} & $0.033$ & - & $0.040$ \\
$\gamma_\text{dl}$ & \(\displaystyle \frac{C_\text{dl}RT\mathcal{A}L/F\bar{i}}{c^\text{max}FL/\bar{i}} \) & \tablewordfrac{capacitance time-scale}{discharge time-scale} & $2.1\times10^{-5}$ & - & $1.7\times10^{-4}$ \\
$\omega_c$ & \(\displaystyle \frac{c^\text{max}M_\text{e}}{\rho^\text{max}}\left(1-\frac{M_\text{w}\bar{V}_\text{e}}{M_\text{e}\bar{V}_\text{w}}\right)\) & \makecell{Diffusive kinematic \\ relationship coefficient} & \multicolumn{3}{c|}{$0.70$}\\
$\omega_i$ & \(\displaystyle \frac{c^\text{max}M_\text{e}}{\rho^\text{max}}\left(t_+^\text{w}+\frac{M_-}{M_\text{e}}\right)\) & \makecell{Migrative kinematic \\ relationship coefficient} & \multicolumn{3}{c|}{$0.41$}\\
$\pi_\text{os}$ & \(\displaystyle \frac{\mu^\text{max}v^\text{rxn}L/d^2}{RTc^\text{max}} \) & \tablewordfrac{viscous pressure scale}{osmotic pressure scale} & \multicolumn{3}{c|}{$3.6\times10^{-5}\mathcal{C}$}\\
\hline
\end{tabular}
\caption{Dimensionless parameters, relative to the C-rate, $\mathcal{C} = 8A_\text{cs}\bar{i}/Q$. $\Cd$ is the diffusional C-rate.}
\label{tab:dimless_params}
\end{table*}

\section{Numerical solution}
\label{sec:sol_num}

The system of equations \eqref{eq:summary} was solved numerically. Code used to solve the model and generate the results below is available publicly on GitHub \cite{valentin_sulzer_2019_2554000}.

To facilitate numerical solution, the model was first manipulated into a form suitable for application of the Finite Volume Method.
Letting $\phi = \Phi_\text{s}-\Phi$ and noting that $i_\text{s} = \mathrm{i}_\text{cell} - i$, one can replace equations \eqref{eq:i} to \eqref{eq:is} with
\bse\label{eq:is_Delta_substitution}
\begin{align}
{\Cd\,{i}} &= \keff\left(\chi\pd{\ln(c)}{x} - \pd{\Phi}{x}\right), \\
\mathrm{i}_\text{cell} - i &= -\iota_\text{s}\pd{}{x}(\phi + \Phi).
\end{align}
\ese
We also eliminate $\partial \Phi/\partial x$ from equation \eqref{eq:is_Delta_substitution} to find $i$ as a functional of $c$ and $\phi$:
\bse\label{eq:numerical_summary}
\begin{equation}\label{eq:i_functional_c_xi}
i = \keff\frac{\chi\pd{\ln(c)}{x} + \pd{\phi}{x} + \mathrm{i}_\text{cell}/\iota_\text{s}}{\Cd + \keff/\iota_\text{s}}.
\end{equation}
The result is a closed system of PDEs for $c$, $\varepsilon$ and $\phi$:
\begin{align}
\pd{}{t}\left(\varepsilon c\right) &= \frac{1}{\Cd}\pd{}{x}\left(\Deff\pd{c}{x}\right) + sj,\\
\pd{\varepsilon}{t} &= -\beta^\text{surf}j,  \\
\pd{\phi}{t} &= \frac{1}{\gamma_{\text{dl}}}\left(j - 2\jecd\sinh\left[\phi - {U}(c)\right]\right)
\end{align}
where $j = \partial i/\partial x$ is given by equation \eqref{eq:i_functional_c_xi} in each electrode and vanishes in the separator,
with boundary conditions
\begin{align}
\pd{c}{x} = i = 0 \quad &\text{ at } x = 0, 1. \\
i = \mathrm{i}_\text{cell} \quad &\text{ at } x = \ell_\text{n}, 1-\ell_\text{p},
\end{align}
and initial conditions derived from \eqref{eq:ICs}.

Equation system \eqref{eq:numerical_summary} is solved by discretising the spatial domain using Finite Volumes, choosing the spatial discretisation to be uniform within each subdomain and as uniform as possible across subdomains %(10 points in the negative electrode, 17 in the separator, and 14 in the positive electrode).
The results are robust to the total number of points chosen.%; 41 were chosen to give good accuracy in a relatively small amount of time.

Having discretised in space, the resulting system of transient ordinary differential equations is solved using \texttt{scipy.integrate} in Python \cite{scipy}. Finally, $\Phi$ is calculated as
\begin{equation}
\Phi = \myint{0}{x}{\left({\chi}\pd{\ln(c)}{x} - \Cd\frac{i}{\keff}\right)}{x} - \left.\phi\right\rvert_{x=0},
\end{equation}
where the final term comes from the fact that $\Phi = -\phi$ at $x=0$. The voltage drop across the electrode pair is computed with
\begin{equation}
V = \left.\Phi_\text{s}\right\rvert_{x=1} = \left.\left(\phi + \Phi\right)\right\rvert_{x=1}.
\end{equation}
\ese
Crucially, the system of partial differential equations \eqref{eq:numerical_summary} is much easier to solve than the differential-algebraic system \eqref{eq:summary}.

\section{Results}
\label{sec:results}
%\tikzsetnextfilename{compare_voltages_numerical}
\begin{figure}
	\centering
%	\if \fullbuild1 \input{figures/calls/voltages_numerical_capacity} \fi
	\includegraphics{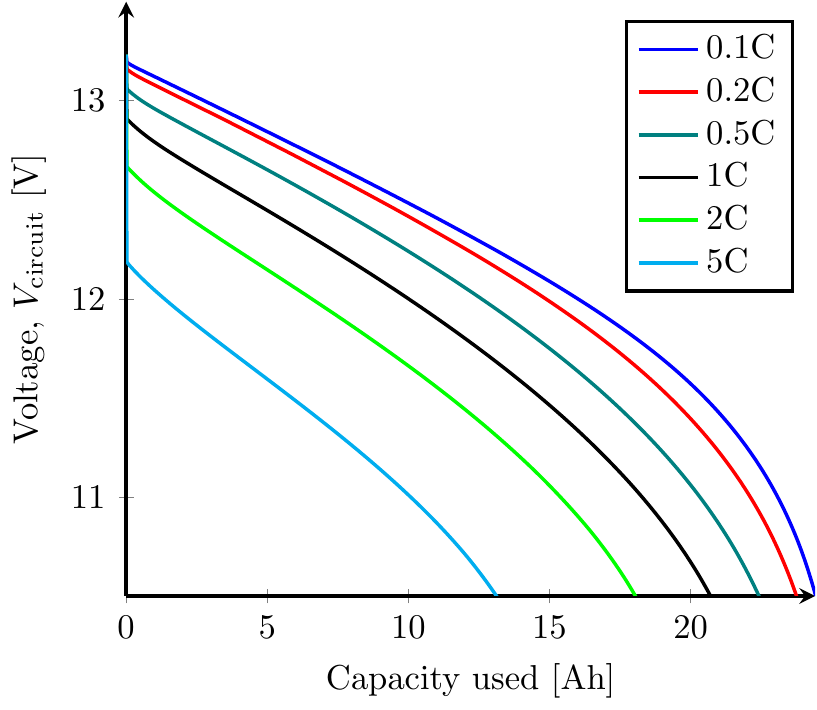}
	\caption{Comparing voltages for a constant-current discharge using the parameters from literature (Table \ref{tab:dim_params}), for a range of C-rates.}
	\label{fig:voltages_numerical}
\end{figure}
\begin{figure}
	\centering
%	\if \fullbuild1 \input{figures/calls/capacities} \fi
	\includegraphics{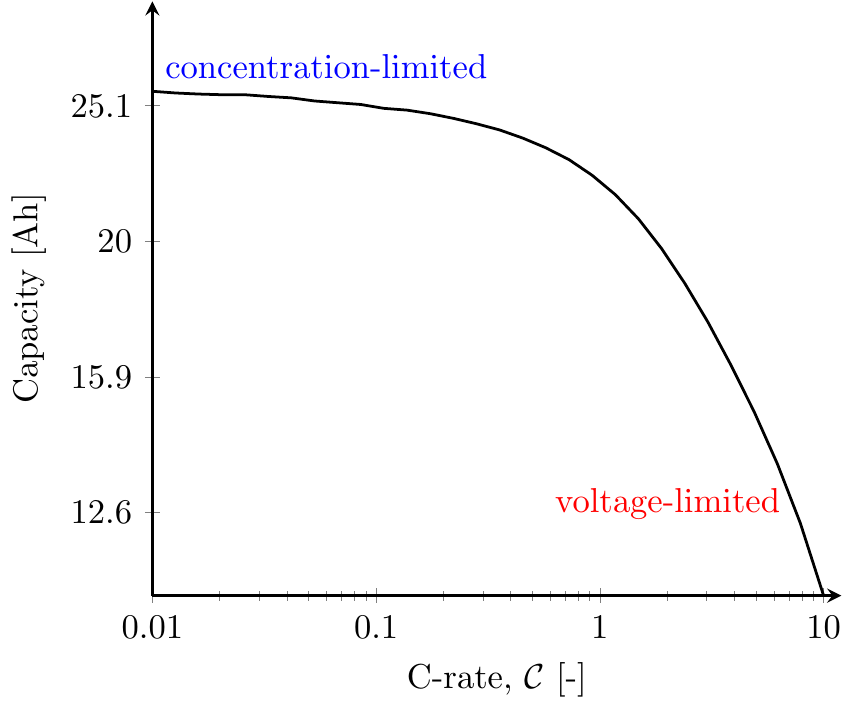}
	\caption{Battery capacity against C-rate, using the parameters from literature (Table \ref{tab:dim_params}).}
	\label{fig:Peukert}
\end{figure}

%\paragraph{Voltage and capacity}
Figure \ref{fig:voltages_numerical} shows the (dimensional) voltage against capacity used for a range of C-rates. As is to be expected, the total capacity available decreases as the C-rate increases.
This dependence is further elucidated by exploring the total capacity of the battery for constant-current discharges across a higher range of C-rates, summarized on Figure \ref{fig:Peukert}.
By Peukert's Law, one would expect the graph to be linear on this log-log axis. However, in this case deviations from linearity occur because there are two distinct capacity-limiting mechanisms. At low C-rates (below 1C), the battery capacity is concentration-limited, with full discharge occurring when the electrolyte concentration reaches zero. At high C-rates, the battery is voltage-limited, because the cut-off voltage of 10.5V is achieved before the bulk concentration falls to zero.

%\tikzsetnextfilename{capacities}

\def \Iinternala {0.1C}
\def \Iinternalb {0.5C}
\def \Iinternalc {2C}
% \tikzsetnextfilename{compare_concentrations_numerical}
\begin{figure*}[t]
	\centering
%	\if \fullbuild1
	% \input{figures/calls/compare_concentrations_numerical}
	%	\fi
	\includegraphics{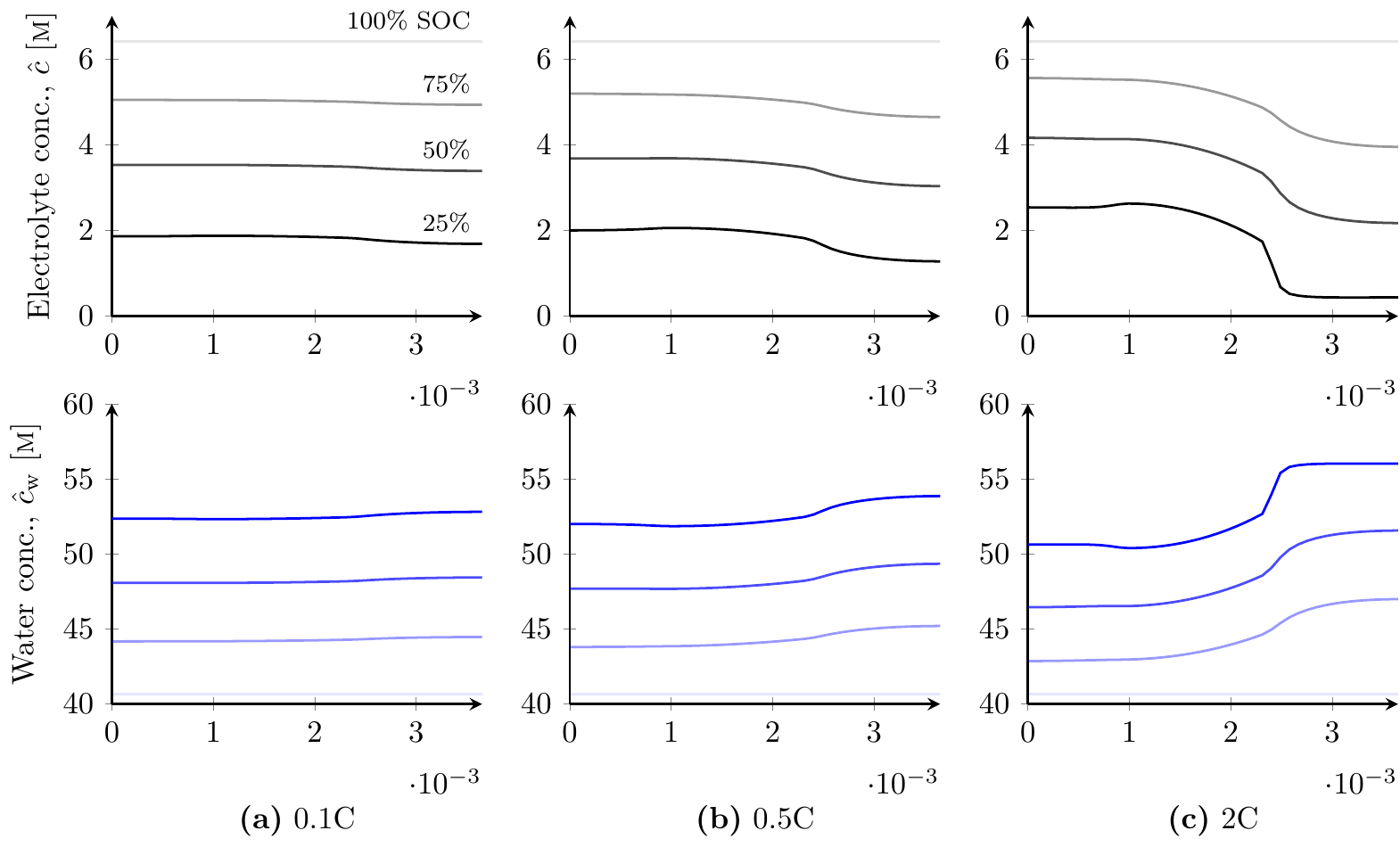}
	\caption{Electrolyte and water concentrations at various States of Charge (SOCs) for a constant-current discharge using the parameters from literature (Table \ref{tab:dim_params}), for a range of C-rates. Opacity increases with decreasing SOC.}
	\label{fig:compare_concentrations_numerical}
\end{figure*}

%\paragraph{Concentration}
The model also allows internal variables to be explored, particularly the local water concentration. Electrolyte and water concentrations are depicted in Figure \ref{fig:compare_concentrations_numerical}. At a very low C-rate of \Iinternala~(Figure \ref{fig:compare_concentrations_numerical}a), both concentrations remain almost uniform throughout the discharge; electrolyte concentration decreases, and water concentration increases proportionally.
At a higher C-rate of \Iinternalb~(Figure \ref{fig:compare_concentrations_numerical}b)
and \Iinternalc~(Figure \ref{fig:compare_concentrations_numerical}c), the concentrations become spatially inhomogeneous, which leads to concentration overpotentials that limit the accessible capacity.

%\tikzsetnextfilename{fits_numerical}
\begin{figure}
	\centering
%	\if \fullbuild1 \input{figures/calls/fits_numerical} \fi
	\includegraphics{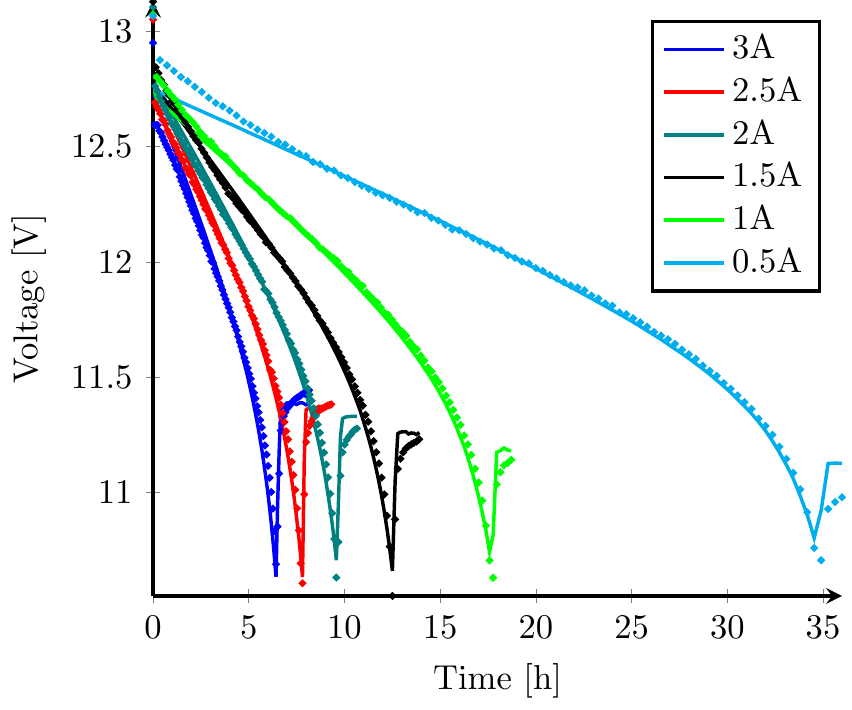}
	\caption{Comparing data (dots) with results from full numerical model (lines) for a range of currents, with parameters fitted using DFO-GN.}% (Tables \ref{tab:dim_params} and \ref{tab:fit_params}).}
	\label{fig:fits}
\end{figure}

%\paragraph{Parameter fitting}
A Derivative-Free Gauss-Newton method~\cite{cartis2017derivative} was also used to fit the model to data from a series of constant-current discharges of a 17 Ah BBOXX Solar Home battery at intervals of 0.5 A from 3 A to 0.5 A (Figure \ref{fig:fits}). Each constant-current discharge is followed by a two-hour rest period during which the current is zero. The fit is good, but this approach is slow since it requires solving the full PDE system at each iteration. A faster approach will be developed in part II.
% \begin{table*}[h]
% \centering
% \begin{tabular}{|c c|c|c|c c c c c|}
% \hline
% $\varepsilon_{\text{n},\text{p}}^\text{max}$ & $\varepsilon_\text{sep}^\text{max}$ & $j_\text{n}^\text{ref}$ & $R_\text{circuit}$ & $q^0_\text{2.5A}$ & $q^0_\text{2A}$ & $q^0_\text{1.5A}$ & $q^0_\text{1A}$ & $q^0_\text{0.5A}$ \\
% \hline
% \input{code/out/tables/fits/dfogn_Numerical_params.txt} \\
% \hline
% \end{tabular}
% \caption{Parameters obtained by fitting the model to experimental data using a Derivative-Free Gausse-Newton method \cite{cartis2017derivative}, with the values in Table \ref{tab:dim_params} as initial data for the parameters.}
% \label{tab:fit_params}
% \end{table*}

\section{Conclusions}
\label{sec:conc}

Three novel phenomena were included in a porous-electrode model for lead-acid batteries.
First, the mass-averaged and volume-averaged velocities of the electrolyte were both considered, the former associated with momentum transport, and the latter with kinematics. Due to density variation in the liquid, and volume changes associated with the electrode reactions, neither of these velocities remains solenoidal after homogenisation.
Second, an extra convective term, associated with the volume-averaged velocity, drives cation transport.
Third, a pressure-diffusion term drives cation transport, and appears also in the MacInnes equation (modified Ohm's law) describing the liquid.
Although these terms are small in magnitude for lead-acid batteries, they could be important for other chemistries where large volume changes occur during charge/discharge, such as lithium-ion batteries with silicon anodes \cite{ bower2011finite,
chon2011real,
sethuraman2011situbiaxial,
sethuraman2010situstress,
sethuraman2010situpotential,
sethuraman2011increased}, or to understand the impedance signature of electromechanical/transport coupling \cite{goyal2017exploring}.
Nondimensionalisation of this model allows us to identify key parameter groupings, and could also easily be extended to other models and chemistries.

Two distinct mechanisms determine the total cell capacity: concentration limitation at low C-rates, and voltage limitation at high C-rates.
In addition to capacity-limitation and voltage-limitation, there may be additional capacity limitations from additional physical effects, such as pore occlusion at very high C-rates \cite{liu2016capacity}.

The model developed in this paper provides physical detail about the electrochemical processes occurring in a lead-acid battery during discharge, but is ultimately too computationally intensive to be used for advanced battery management systems.
In part II, asymptotic analysis is used to derive three simplified models valid at low-to-moderate discharge rates. These can be solved much faster than the detailed model developed here, while giving additional physical insights.

\section*{Acknowledgements}
\label{sec: ack}

This publication is based on work supported by the EPSRC center For Doctoral Training in Industrially Focused Mathematical Modelling (EP/L015803/1) in collaboration with BBOXX. JC, CP, DH and CM acknowledge funding from the Faraday Institution (EP/S003053/1).

\section*{List of symbols}

\noindent\textbf{Variables}
\begin{description}[leftmargin=!, labelwidth=1cm, font=\normalfont]
  \item[$c$] concentration \hfill mol m$^{-3}$
  \item[$\varepsilon$] porosity \hfill -
  \item[$j$] interfacial current density \hfill A m$^{-2}$
  \item[$\bm{i}$] current density (3D) \hfill A m$^{-2}$
  \item[$i$] current density in $x$-direction \hfill A m$^{-2}$
  \item[$\bm{v}$] velocity (3D) \hfill m s$^{-1}$
  \item[$v$] velocity in $x$-direction \hfill m s$^{-1}$
  \item[$\bm{N}$] ion flux (3D) \hfill mol m$^{-2}$ s$^{-1}$
  \item[$p$] pressure \hfill Pa
  \item[$\Phi$] potential \hfill V
\end{description}

\noindent\textbf{Subscripts}
\begin{description}[leftmargin=!, labelwidth=1cm, font=\normalfont]
  \item[n] in negative electrode
  \item[sep] in separator
  \item[p] in positive electrode
  \item[$+$] of cations
  \item[$-$] of anions
  \item[w] of solvent (water)
  \item[e] of electrolyte
  \item[s] of solid (electrodes)
\end{description}

\noindent\textbf{Superscripts}
\begin{description}[leftmargin=!, labelwidth=1cm, font=\normalfont]
  \item[$0$] initial
  \item[max] maximum
  \item[eff] effective
  \item[surf] surface
  \item[$\square$] convective
\end{description}

\noindent\textbf{Accents}
\begin{description}[leftmargin=!, labelwidth=1cm, font=\normalfont]
  \item[$\di{}$] dimensional
\end{description}

\begin{appendix}

\section{Parameters}
\label{app:params}
\setcounter{table}{0}

The dimensional parameters are given in Table \ref{tab:dim_params}, and the concentration dependences of coefficients are laid out in Table \ref{tab:dim_functions}.
Formulae for open-circuit potentials were obtained empirically by Bode \cite{bode1977lead}. Note that these could be written as
\bse\label{eq:OCPs_split}
\begin{align}
\di{U}\ocpn(\di{c}) &= U^\ominus\ocpn + \frac{RT}{F}{U}\ocpn({c}), \\
\di{U}\ocpp(\di{c}) &= U^\ominus\ocpp + \frac{RT}{F}{U}\ocpp({c}),
\end{align}
\ese
to reflect their relation to the Nernst equation (e.g.~\cite{treptow2002lead}), with $U^\ominus\ocpn = -0.295$ and $U^\ominus\ocpp = 1.628$.

\renewcommand{\arraystretch}{1}
\begin{table*}[t]
\centering
\begin{tabular}{|c|c c c|c|c|}
\hline
\multirow{2}{*}{Parameter} & \multicolumn{3}{c|}{Value} & \multirow{2}{*}{Units} & \multirow{2}{*}{Reference} \\
\cline{2-4}
& n & sep & p & & \\
\hline
$l$ & $0.9\times10^{-3}$ & $1.5\times10^{-3}$ & $1.25\times10^{-3}$ & m & Private communication \\
$\hat{\varepsilon}^\text{max}$ & $0.53$ & $0.92$ & $0.57$ & - & ~\cite{srinivasan2003mathematical}\\
$H$ & \multicolumn{3}{c|}{$11.4\times10^{-2}$} & m & Measured \\
$W$ & \multicolumn{3}{c|}{$6.5\times10^{-2}$} & m & Measured \\
$s_{+}$ & $-1$ & - & $-3$ & - & \eqref{eq:halfreactions} \\
$s_{-}$ & $1$ & - & $-1$ & - & \eqref{eq:halfreactions} \\
$s_\text{w}$ & $0$ & - & $2$ & - & \eqref{eq:halfreactions} \\
$n_\text{e}$ & $2$ & - & $2$ & - & \eqref{eq:halfreactions} \\
$\bar{V}_\text{w}$ & \multicolumn{3}{c|}{$1.75\times 10^{-5}$} & m$^3$ mol$^{-1}$ &~\cite{bernardi1995mathematical} \\
$\bar{V}_\text{e}$ & \multicolumn{3}{c|}{$4.50\times 10^{-5}$} & m$^3$ mol$^{-1}$ &~\cite{bernardi1995mathematical}\\
$\bar{V}_+$ & \multicolumn{3}{c|}{$1.35\times 10^{-5}$} & m$^3$ mol$^{-1}$ &~\cite{boovaragavan2009mathematical}\\
$\bar{V}_-$ & \multicolumn{3}{c|}{$3.15\times 10^{-5}$} & m$^3$ mol$^{-1}$ &~\cite{boovaragavan2009mathematical}\\
$\bar{V}\ocpn$ & \multicolumn{3}{c|}{$1.83\times 10^{-5}$} & m$^3$ mol$^{-1}$ &~\cite{lide1992crc}\\
$\bar{V}\ocpp$ & \multicolumn{3}{c|}{$2.55\times 10^{-5}$} & m$^3$ mol$^{-1}$ &~\cite{lide1992crc}\\
$\bar{V}_{\ce{PbSO_4}}$ & \multicolumn{3}{c|}{$4.82\times 10^{-5}$} & m$^3$ mol$^{-1}$ &~\cite{lide1992crc}\\
$M_\text{w}$ & \multicolumn{3}{c|}{$1.8\times 10^{-2}$} & kg mol$^{-1}$ &~\cite{lide1992crc}\\
$M_+$ & \multicolumn{3}{c|}{$0.1\times 10^{-2}$} & kg mol$^{-1}$ &~\cite{lide1992crc}\\
$M_-$ & \multicolumn{3}{c|}{$9.7\times 10^{-2}$} & kg mol$^{-1}$ &~\cite{lide1992crc}\\
$F$ & \multicolumn{3}{c|}{$96485$} & C mol$^{-1}$ & Faraday constant \\
$R$ & \multicolumn{3}{c|}{$8.314$} & J mol$^{-1}$ K$^{-1}$ & Ideal gas constant \\
$T$ & \multicolumn{3}{c|}{$298.15$} & K & Room temperature \\
$t^\text{w}_+$ & \multicolumn{3}{c|}{$0.72$} & - &~\cite{chapman1968compilation, gu1987mathematical} \\
$\sigma$ & $4.8\times10^6$ & - & $8\times10^4$ & S m$^{-1}$ &~\cite{gu1997numerical}\\
$j^\text{ref}$ & $8\times10^{-2}$ & - & $6\times10^{-3}$ & A m$^{-2}$ &~\cite{tiedemann1979battery} (reported by~\cite{gu1997numerical}) \\
$c^\text{max}$ & \multicolumn{3}{c|}{$5.6\times10^{3}$} & mol m$^{-3}$ & Private communication \\
$\mathcal{A}$ & $2.6\times10^6$ & - & $2.05\times10^7$ & m$^{-1}$ &~\cite{tiedemann1979battery} (reported by~\cite{gu1997numerical}) \\
$d$ & $10^{-7}$ & - & $10^{-7}$ & m &~\cite{gu1997numerical} \\
$C_\text{dl}$ & $0.2$ & - & $0.2$ & F m$^{-2}$ &~\cite{srinivasan2003mathematical}\\
$q^0$ & \multicolumn{3}{c|}{$1$} & - & Full initial SOC \\
$Q$ & \multicolumn{3}{c|}{$17$} & Ah & Manufacturer-specified \\
\hline
\end{tabular}
\caption{Dimensional parameters from the literature. Parameters with several values indicate different values in negative electrode (n), separator (s) and positive electrode (p).}
\label{tab:dim_params}
\end{table*}

\begin{table*}[t]
\centering
\begin{tabular}{|c|c|c|c|c|}
\hline
Function & Formula & \makecell[c]{Value at \\ $\di{c}=c^\text{max}$}  & Units & Reference \\
\hline
$\di{D}$ & $(1.75+2.6\times10^{-4}\di{c})\times10^{-9}$ & $3.02\times10^{-9}$ &m$^2$ s$^{-1}$& (\textdagger) \\
$\di{\chi}$ & $0.49 + 4.1\times10^{-4}\di{c}$ & $2.8$ & - &~\cite{chapman1968compilation, pitzer1977thermodynamics} (*)\\
$\di{\kappa}$ & \makecell[l]{$\di{c}\exp\left(6.23 - 1.34\times10^{-4}\di{c}\right.$\\\hspace{2cm}$\left. - 1.61\times10^{-8}\di{c}^2\right)\times10^{-4}$} & $77$ & S m$^{-1}$ & (\textdagger) \\
\hline
$\di{\rho}$ & $M_\text{w}/\bar{V}_\text{w}\left(1+\left(M_e\bar{V}_\text{w}/M_\text{w} - \bar{V}_e\right)\di{c}\right)$ &	$1.32\times10^3$ & kg m$^{-3}$& Linear function of $\di{c}$~\cite{newman2012electrochemical} \\
$\di{\mathcal{K}}$ & $\di{\varepsilon}^3d^2/180(1-\di{\varepsilon})^2$ & - & m$^2$ &~\cite{gu1997numerical} \\
$\di{\mu}$ & $0.89\times10^{-3} + 1.11\times10^{-7}\di{c} + 3.29\times10^{-11}\di{c}^2$ & $2.5\times10^{-3}$ & Pa s &~\cite{chapman1968compilation} (*) \\
$\di{c}_\text{w}$ & $(1-\di{c}\bar{V}_\text{e})/\bar{V}_\text{w}$ &$4.27\times10^4$ & mol m$^{-3}$ & \eqref{eq:cw}\\
\hline
$\di{j}_0$ & $j^{\text{ref}} \left( \frac{\di{c} }{c^{\text{ref}} } \right)^{ \left| \frac{s_+}{n_{\text{e}}} \right| + \left| \frac{s_-}{n_{\text{e}}} \right| } \left( \frac{\di{c}_{\text{w}} }{c_{\text{w}}^{\text{ref}} } \right)^{\left| \frac{s_\text{w}}{n_{\text{e}}} \right| }$ & $j^{\text{ref}}$ & A m$^{-2}$ & Approach of~\cite{liu2016capacity} \\
$\di{U}\ocpn$ & \makecell[l]{$-0.295 - 0.074\log m- 0.030\log^2m$ \\\hspace{.9cm}$- 0.031\log^3m - 0.012\log^4m$ (\textdaggerdbl)} & $-0.41$ & V &~\cite{bode1977lead} \\
$\di{U}\ocpp$ & \makecell[l]{$1.628 + 0.074\log m+ 0.033\log^2m$ \\\hspace{.9cm}$+ 0.043\log^3m + 0.022\log^4m$ (\textdaggerdbl)} & $1.76$ & V &~\cite{bode1977lead} \\
\hline
\end{tabular}
\caption{Dimensional functions of concentration, $\di{c}$ (measured in mol/m$^3$) and $\di{\varepsilon}$ (dimensionless). (\textdagger) Empirical formulae are given by~\cite{gu1997numerical}, citing~\cite{tiedemann1979battery}, and agree with data of~\cite{chapman1968compilation}. (*) Our fit to data of given reference(s). (\textdaggerdbl) $m(\di{c}) = \di{c}\bar{V}_\text{w}/[(1-\di{c}\bar{V}_\text{e})M_\text{w}]$.}
\label{tab:dim_functions}
\end{table*}

\paragraph{Consistent initial conditions} For consistent initial conditions, we take `starting at $x\%$ SOC' to mean an internally equilibrated (i.e. spatially homogeneous) initial state at uniform concentration corresponding to this charge. The following determines initial conditions for $\di{c}$ and $\di{\varepsilon}$ consistent with this choice.

Considering a one-dimensional model and integrating \eqref{eq:dim_saltbal} in $\di{x}$ across the whole electrode pair and using the no-flux conditions \eqref{eq:no-flux_sides}, and the integral condition \eqref{eq:dim_j_BC},
\begin{multline}\label{eq:inits_find_qmax}
{\myint{0}{L}{\left(\di{\varepsilon} \di{c}\right)}{\di{x}}} = A_\text{cs}\left(l_\text{n}\varepsilon_\text{n}^\text{max} + l_\text{sep}\varepsilon_\text{sep}^\text{max} + l_\text{p}\varepsilon_\text{p}^\text{max}\right) c^\text{max} \\
+ \frac{(s_\text{n}-s_\text{p})q^\text{max}(1-q)}{8F}.
\end{multline}
Both sides of \eqref{eq:inits_find_qmax} should be zero when $q=0$; hence choose
\begin{equation}
q^\text{max} = \frac{8FA_\text{cs}\left(l_\text{n}\varepsilon_\text{n}^\text{max} + l_\text{sep}\varepsilon_\text{sep}^\text{max} + l_\text{p}\varepsilon_\text{p}^\text{max}\right) c^\text{max}}{s_\text{p}-s_\text{n}},
\end{equation}
which, with the parameter values in Table \ref{tab:dim_params}, gives a maximum capacity of $26.1$ Ah for the battery. This compares favourably to the stated battery capacity of $17$ Ah.

Now, integrating \eqref{eq:dim_depsdt} in $\di{x}$ across the whole negative electrode and using the integral condition \eqref{eq:dim_j_BC},
\begin{equation}\label{eq:inits_find_epsDelta}
{\myint{0}{l_\text{n}}{\di{\varepsilon}_\text{n}}{\di{x}}} = A_\text{cs}l_\text{n}\varepsilon_\text{n}^\text{max} + \frac{\Delta\bar{V}^\text{surf}_\text{n}q^\text{max}(1-q)}{16F}.
\end{equation}
Then, assuming that $\varepsilon^0_\text{n}$ is spatially uniform,
\begin{equation}
\di{\varepsilon}^0_\text{n} = \di{\varepsilon}_\text{n}^\text{max} - \di{\varepsilon}^\Delta_\text{n}(1-q^0),
\quad \text{where} \quad
\varepsilon^\Delta_\text{n} = \frac{\Delta\bar{V}^\text{surf}_\text{n}q^\text{max}}{16FA_\text{cs}l_\text{n}}.
\end{equation}
Similarly,
\begin{equation}
\di{\varepsilon}^0_\text{p} = \di{\varepsilon}_\text{p}^\text{max} - \di{\varepsilon}^\Delta_\text{p}(1-q^0),
\quad \text{where} \quad
\varepsilon^\Delta_\text{p} = \frac{\Delta\bar{V}^\text{surf}_\text{p} q^\text{max}}{16FA_\text{cs}l_\text{p}}.
\end{equation}

\section{Velocity}
\label{app:velocity}

Integrating \eqref{eq:vbox} for the volume-averaged velocity from $x=0$ to $x=1$ and using the integral condition \eqref{eq:dim_j_BC} for the interfacial current density and \eqref{eq:BCs_collectors} for $v^\square$ at $x=0$, one finds that
\begin{equation}
\left.v^\square\right\rvert_{x=1} = \left(\beta^\text{surf}_\text{n} + \beta^\text{liq}_\text{n} + \beta^\text{surf}_\text{p} + \beta^\text{liq}_\text{p}\right)i_\text{cell},
\end{equation}
which contradicts \eqref{eq:BCs_collectors} for $v^\square$ at $x=1$ since the sum of the volume changes across the whole cell is non-zero. It follows that the model can have no exact solution in a one-dimensional setting. This can be resolved by considering a multi-dimensional problem with a free electrolyte surface; this solution is beyond the present scope. %For the present purposes, we take the limit in which $\left(\beta^\text{surf} + \beta^\text{liq}\right)$ is zero in each electrode.
%Then $v^\square=0$ everywhere, and \eqref{eq:Darcy} and \eqref{eq:kinematic} for $p$ and $v$ decouple from the other equations: having found $c$ and $i$, one could easily determine $v$ and hence $p$. These solutions are not presented in this paper.

\end{appendix}
\bibliographystyle{unsrt}

\end{document}